\def\ii{\'{\i}}

\def\beg{\begin{equation}} 
\def\begl{\begin{equation}\label} 
\def\fim{\end{equation}} 
 
\documentstyle[preprint,aps]{revtex} 
\tightenlines

\begin{document} 
 
\title{Quasi-stationary distributions for the Domany-Kinzel stochastic  
cellular automaton} 
 
\author{	A.P.F. Atman \thanks{atman@fisica.ufmg.br} and
	Ronald Dickman \thanks{dickman@fisica.ufmg.br}}
 
\address{Departamento de F\ii sica, Instituto de Ci\^encias Exatas,\\ 
Universidade Federal de Minas Gerais, C. P. 702\\  
30123-970, Belo Horizonte, MG - Brazil} 
\date{\today} 
\maketitle 
 
\begin{abstract} 
We construct the {\it quasi-stationary} (QS) 
probability distribution for the Domany-Kinzel  
stochastic cellular automaton (DKCA), 
a discrete-time Markov process with an absorbing state. 
QS distributions are derived 
at both the one- and two-site levels.  
We characterize the distribuitions by their mean, and various 
moment ratios, and analyze the lifetime of the QS state, and the
relaxation time to attain this state.
Of particular interest are the scaling properties of the QS state
along the critical line separating the active and absorbing phases.  
These exhibit a high degree of similarity to the contact process and the 
Malthus-Verhulst process (the closest continuous-time
analogs of the DKCA), which extends to the scaling form of
the QS distribution.
\vspace{1em}

\end{abstract} 
 
\newpage
\section{Introduction} 
The Domany-Kinzel stochastic cellular automaton (DKCA) \cite{dk} is  
a Markov process that exhibits a  
phase transition from an active state to an absorbing one.  
Stochastic processes with an absorbing state arise  
frequently in statistical physics \cite{kampen}, and are currently of great  
interest in connection with self-organized criticality \cite{mundick} and  
nonequilibrium critical phenomena \cite{marro,hynri2}. Many studies of the  
DKCA and related probabilistic cellular automata (PCA) have been
published, using deterministic mean-field equations \cite{tome1,bagnoli,guto},  
Monte Carlo simulations \cite{kinzel,martins,alemao,hynri,atman} and  
renormalization group (RG) analyses \cite{tomemario,bagnoli2,kemper}.
 
While the mean-field (MF) description of the DKCA admits (for 
appropriate parameter values) an active stationary state, 
for finite system sizes the model always ends up in
the absorbing state, due to fluctuations. 
MF theories ignore such fluctuations, and so are incapable of
treating finite systems.  But, since simulations and other
numerical methods typically study finite systems, it is of
interest to develop approximate theoretical descriptions
that account for finite system size.
A natural way to study finite systems with an absorbing state is via the
{\it quasi-stationary distribution}, which, when it exists, describes the
asymptotic properties conditioned on survival
\cite{yaglom,ferrari,nasell}.
Recently, 
mean-field-like methods were developed for studying the
quasi-stationary state of finite systems with an  
absorbing state \cite{dickvid}.  
The quasi-stationary properties converge to the
true stationary properties in the infinite-size limit.
(Indeed, this provides the rationale for studying the ``stationary" behavior of
absorbing-state models in simulations, which of necessity treat finite systems.)
In Ref. \cite{dickvid}, quasi-stationary distributions for various continuous-time Markov
processes are constructed, in particular, for the contact process (CP) and the
closely related Malthus-Verhulst process (MVP).  In the case of the CP, both
one- and two-site approximations are derived.  In  this work, we extend the
analysis to {\it discrete-time} processes, using the DKCA as an interesting
example, closely related to the CP.

The remainder of this paper is structured as follows. In Sec. II we 
review the one- and two-site MF approximations for the DKCA.  In  
Sec. III, we construct the quasi-stationary (QS) probability 
distribution at the site level. 
The QS distribution  
at the pair level is discussed in section IV, while Sec. V  
presents our conclusions.
 
\section{Site and pair MF approximations}

In this section we review the definition of the DKCA and its mean-field
description at the one- and two-site levels \cite{tome1,guto}.
The DKCA is a discrete-time Markov
process (all sites are updated simultaneously),
whose configuration is given by a set of stochastic 
variables $\{ \sigma_i \}$ ($\sigma_i = 0$ or 1),  
defined at sites $i$ and times $t\!=\!0$, 1, 2,...,
such that $t+i$ is even. 
Let $\sigma$
represent the configuration at time $t$,
and $P_t(\sigma)$ the probability distribution in configuration space.
The evolution of the latter is governed by  
\beg\label{um} 
P_{t+1}(\sigma) = \sum_{\sigma'}\omega(\sigma|\sigma')P_t(\sigma') 
\fim 
where $\omega(\sigma|\sigma')$ denotes the probability of the 
transition $\sigma' \to \sigma$, and enjoys the properties 
$ \omega(\sigma|\sigma') \ge 0$ and
$\sum_{\sigma}\omega(\sigma|\sigma') = 1$. 
The transition probability for the DKCA is a product of factors associated
with each site: 
\beg\label{quatro} 
\omega(\sigma|\sigma') = \prod_{i=1}^{L} w_i(\sigma_i|\sigma')~~, 
\fim 
where $w_i(\sigma_i|\sigma') $ is the conditional probability 
for site $i$ to be in state $\sigma_i$ at time $t+1$,
given configuration $\sigma'$ at time $t$.  
The probabilities $w_i(\sigma_i|\sigma')$ are translation-invariant
and in fact depend only on the variables $\sigma_{i-1}'$ and $\sigma_{i+1}'$
at the previous time step:
\beg\label{cinco} 
w_i(\sigma_i|\sigma') = w_{DK}(\sigma_i|\sigma'_{i-1}, \sigma'_{i+1})~. 
\fim 
The above relations, with the transition probabilities given in Table I, define
the DKCA.  Noting that the transition (00) $\to $ (1) is prohibited, 
we see that the configuration $\sigma_i \!=\! 0$, $\forall i$ is {\it absorbing}.

Of interest are the $n$-site marginal probabilities.  The evolution of
the one-site distribution 
$P_t(\sigma_i) \equiv \sum_{\sigma_j, j\neq i} P_t(\sigma)$, is given by,
\beg\label{seis} 
P_{t+1}(\sigma_i) = \sum_{\sigma'_{i-1}}\sum_{\sigma'_{i+1}} 
w_{DK}(\sigma_i|\sigma'_{i-1}, \sigma'_{i+1}) 
 P_t(\sigma'_{i-1}, \sigma'_{i+1})~,
\fim 
where $ P_t(\sigma'_{i-1}, \sigma'_{i+1})$ is the marginal distribution
for a pair of nearest-neighbor sites.  The evolution of the latter is
coupled to the three-site probability, so:
\beg\label{dez} 
P_{t+1}(\sigma_{i-1},\sigma_{i+1}) =  
\sum_{\sigma'_{i-2}}\sum_{\sigma'_{i}}\sum_{\sigma'_{i+2}}  
w_{DK}(\sigma_{i-1}|\sigma'_{i-2}, \sigma'_{i})
w_{DK}(\sigma_{i+1}|\sigma'_{i},\sigma'_{i+2})  
P_t(\sigma'_{i-2}, \sigma'_{i},\sigma'_{i+2})~. 
\fim 
Evidently we have an infinite hierarchy of equations.  In the $n$-site
approximation the hierarchy is truncated by estimating the
$(n\!+\!1)$-site probabilitites on the basis of those for $n$ sites.

The simplest case is the one-site approximation \cite{tome1,guto},
in which $ P_t(\sigma'_{i-1}, \sigma'_{i+1})$ 
is factored into a product of one-site
probabilities.  This yields the recurrence relation,  
\beg\label{xum} 
\rho_{t+1} = \rho_t \left[ 2 p_1 - ( 2 p_1 \!-\! p_2) \rho_t \right]~,
\fim 
where $\rho_t \equiv P_t(1)$ is the density of active sites
(the order parameter for the DKCA).
Eq. (\ref{xum}) admits two stationary solutions, corresponding 
to the possible DKCA phases: {\it absorbing} 
($\rho = 0$), and {\it active},  
in which, for $p_1 > 1/2$,
\begl{x} 
\rho = \frac{2p_1 - 1}{2p_1 - p_2}~. 
\fim 
Thus the critical line at the site level is $p_{1c} = 1/2$. 
 
In the pair approximation \cite{tome1} the three-site probability is written
in terms of the two-site quantity, using the conditional
probability:
\begl{umpar} 
P_t(\sigma_{i-2}, \sigma_{i}, \sigma_{i+2}) \simeq  
\frac{P_t(\sigma_{i-2}, \sigma_{i}) P_t(\sigma_{i} \sigma_{i+2})}
{P_t(\sigma_{i})}~. 
\fim 
(The one-site probabilities are given by
$P_t(\sigma_i) = \sum_{\sigma_{i+2}} P_t(\sigma_i,\sigma_{i+2})$).
Letting $z_t \equiv P_t(1,1)$, and
noting that $P_t(1,0) \equiv k_t = \rho_t - z_t$, 
we have the relations,
\begl{onzez} 
z_{t+1}= \frac{1}{\rho_t} \left[ p_2 z_t + p_1k_t \right]^2
+ p_1^2 \frac{k_t^2}{1-\rho_t}~, 
\fim 
and  
\begin{equation}
k_{t+1} = 
\frac{1}{\rho_t} (p_2 z_t \!+\! p_1 k_t) (q_2 z_t \!+\! q_1 k_t)
+ \frac{p_1 k_t}{1-\rho_t}(q_1 k_t + v_t)~, 
\label{kdens} 
\end{equation} 
where $q_i \! \equiv \! 1 \!-\! p_i$,
while for $v_t \equiv P_t(0,0)$ we have: 
\begin{equation} 
v_{t+1} = \frac{1}{\rho_t} \left[ q_2 z_t \!+\! q_1 k_t \right]^2
+ \frac{1}{1-\rho_t} \left[q_1 k_t + v_t \right]^2~. 
\label{vdens}
\end{equation} 
In the active stationary state, these relations imply,
\begl{oitz} 
z = \frac{1 - 2p_1}{p_2 - 2p_1}~\rho ~,
\fim 
which leads to the stationary active-site density,
\begl{xpar} 
\rho = \frac{p_2 (p_1 - 1)^2 + p_1(3p_1 - 2)}{(2p_1 - 1)(2p_1 - p_2)}~. 
\fim 
In this approximation, the critical line in the $(p_1,p_2)$ plane is:
\begl{xtr} 
p_2 = \frac{p_1(2 - 3p_1)}{(1-p_1)^2}~. 
\fim 
The phase diagram for the DKCA in the one- and two-site approximations 
is compared with simulation results \cite{atman2}
in Fig. 1. 
 
\section{Quasi-stationary probability distributions} 
 
\subsection{Method}

Since the approximations discussed in the previous section
effectively consider the
$L \to \infty$ limit, the densities ($\rho$, $z$, etc.), are in fact
{\it deterministic} variables.
Our goal in this paper is to construct reduced 
{\it stochastic} descriptions
of a finite system, in a manner analogous to that employed in
deriving $n$-site approximations, and to determine the associated
quasi-stationary properties.  In this section we study the problem at the
one-site level.  Consider the DKCA on a ring of $L$ sites.
At the one-site level, the state of the system is specified by
$N_t$, the number of active sites at time $t$.
Let $p_t(N)$ ($N= 0,...,L$) be the probability to have
exactly $N$ active sites at time $t$.
The probability vector
$p_t= [p_t(0),p_t(1), ..., p_t(L)]$, satisfies
\begl{distr} 
p_t (N)= \sum_{N'} W(N|N') p_{t-1}(N')~,
\fim 
where $W$ is the transition matrix, with
$W(N|N')$ representing the probability to have $N$ 
active sites at time $t$, given $N'$ at time $t \!-\! 1$. 

At the one-site level, the state each site is treated as an independent event.  
Given $N'$ active sites at time $t$, our best estimate for
the probability $x$ of a given site to be active at the next
time step is (see Eq. (\ref{xum}):
\beg\label{xuma} 
x = y \left[ 2 p_1 - ( 2 p_1 \!-\! p_2) y \right]~,
\fim 
where $y \!=\! N'/L$.
Thus the transition probabilitites in the 1-site approximation are:
\begl{ptrans} 
W(N|N') = \frac{L!}{(L\!-\!N)! N!}~ x^N~ (1 \!-\! x)^{L-N}~. 
\fim 
(Here we suppose that all configurations with the same number of
active sites are equally probable, since there is no reason to
prefer one such configuration over another at this level of
analysis.)
The one-site distribution $p_t(N)$ is therefore a 
superposition of binomial distributions with means $x = 0, 1,...,L$,
the weight of a given distribution depending only on the mean
population at the previous step.  

Since (for $L$ finite) the probability distribution 
will always evolve to the
absorbing state, $p(N) = \delta_{N,0}$, it is of interest to study the
quasi-stationary distribution $\overline{p}(N)$, defined as follows.
We suppose that as $t \to \infty$, the probability distribution,
{\it conditioned on survival}, attains a time-independent form.
This means for long times
\begin{equation}
p_t(N) = A_t \delta_{N,0} + S_t \overline{p}(N),
\label{qss}
\end{equation}
where the only time dependence lies in $A_t$ and $S_t$.
Since the QS distribution $\overline{p}(N)$ is conditioned on survival,
$\overline{p}(0) \equiv 0$.  Adopting the normalization 
\begin{equation}
\sum_{N=1}^L \overline{p}(N) = 1,
\label{norm}
\end{equation} 
$S_t$ in Eq. (\ref{qss}) represents the survival probability and
$A_t = 1 \!-\! S_t$ the probability to have fallen into the absorbing state.
The QS hypothesis is verified numerically below.  
Evolving the distribution in
Eq. (\ref{qss}) to the next time step, we have
\begin{eqnarray}
\nonumber
p_{t+1}(N) &=& A_t \delta_{N,0} + S_t \sum_{N'=0}^L W(N|N') \overline{p}(N')
\\
&=& A_{t+1} \delta_{N,0} + S_{t+1} \overline{p}(N),
\label{qss1}
\end{eqnarray}
which implies that $S_{t+1} = \alpha S_t$, where
\begl{alfa} 
\alpha = 1 - \sum_{N=1}^{L} W(0|N) \overline{p}(N)~. 
\fim 
Restricted to the states $1,...,N$, $\overline{p}$ is
an eigenvector of matrix $W$ with eigenvalue $\alpha$.
The lifetime $\tau$ of the QS state is:
\begl{tau} 
\tau = -\frac{1}{\ln \alpha}~. 
\fim 

One method for generating the QS distribution is via iteration of the
evolution equation, Eq. (\ref{distr}), until the distribution
$q_t(N) \equiv p_t(N)/S_t$ (for $N \!=\! 1, 2,...,L$), 
attains a time-independent form.  We refer to this as the {\it direct} method.
An alternative method \cite{intme} is based on writing the evolution in the
form:
\begl{mes} 
\Delta p_t(N) \equiv p_{t+1}(N) - p_t(N) = -w(N) p_t(N) + r_t(N), 
\fim 
where $r_t(N) = \sum_{N' \neq N} W(N|N') p_t(N')$ and 
$w(N) = \sum_{N' \neq N} W(N'|N)$.  Inserting the normalized QS distribution
$\overline{p}(N)$ in the r.h.s. of the above relation, we have
\begin{equation}
(\alpha \!-\! 1) \overline{p}(N) = -w(N)\overline{p}(N) + \overline{r} (N),
\label{qss2}
\end{equation}
where $\overline{r}(N) = \sum_{N'} W(N|N') \overline{p}(N')$.  Noting that
$1 \!-\! \alpha = \overline{r}(0)$, this may be written in the form

\begl{qs} 
\overline{p}(N) = \frac{\overline{r}(N)}{w(N) - \overline{r}(0)}~, \;\;\;\;
N \ge 1~. 
\fim 
This relation suggests the following iterative scheme: 
\begl{iter} 
p'(N) = ap(N) + (1\!-\!a) \frac{r(N)}{w(N) - r(0)}~, 
\fim 
where $a$ is a parameter and $r(N)$ is evaluated using the 
distribution $p(N)$. 
At each iteration the new distribution $p'$ must be normalized.
In this way, we can 
construct the quasi-stationary state from any initial 
initial distribution $p(N)$ that is nonnegative and normalized. 
We call this the {\it iterative} scheme.
As discussed in Ref. \cite{intme}, good convergence is obtained 
for $a\simeq 0$.

\subsection{Results} 

We have constructed the QS distribution for the DKCA at the one-site level,
using both the direct and iterative schemes. 
In Fig. 2 we show  
the time evolution of the probability distribution (conditioned on survival)
at a point on the critical line.  It is evident that the distribution
reaches a quasi-stationary form after about 100 time steps.
Fig. 3 shows the evolution of the mean population $\langle N \rangle$,
the moment ratio $m = \langle N^2 \rangle/\langle N \rangle^2$,
(both conditioned on survival),
and the decay rate $\gamma = r_t(0)/S_t$, 
to their quasi-stationary values, for the same parameters as in Fig. 2.  
(Note that $\gamma$ is the transition rate into the absorbing state.)

Relaxation to the
QS state appears to consist of two stages: an initial transient, which
depends strongly on the initial condition, and a long-time,
exponential approach to the final values.  The mean population, for
example, follows $|\langle N \rangle - \langle N \rangle_{QS}|
\sim e^{-t/\tau_R}$.  (For the data in Fig. 3, $\tau_R \sim 9$,
while the lifetime of the QS state is about 16.) 
We find that $m$ and $\gamma$ {\it relax at the
same rate} as $\langle N \rangle$, and that this relaxation time
is {\it independent of the initial distribution}.  Thus the asymptotic
relaxation to the QS state is governed by a relaxation time $\tau_R$
that appears to depend only on the parameters $p_1$ and $p_2$ and on the
system size $L$.

The distribution may be further characterized by its 
skewness, $S$, and kurtosis, $K$, 
defined through the relations: 
\begl{S} 
S = \frac{\kappa_3}{\kappa_2^{3/2}}~, 
\fim 
and
\begl{K} 
K = \frac{\kappa_4}{\kappa_2^2},
\fim 
where $\kappa_n $ is the $n^{th}$ cumulant of the
distribution \cite{kampen}. 
For the Gaussian distribution, both skewness and kurtosis are null ($S=K=0$). 
The evolution of the skewness and kurtosis is also shown in Fig. 3.
Fig. 4 shows the quasi-stationary distribution at several points in  
parameter space. We observe that in the frozen phase the distribution 
collapses to $N\!=\!1$ while in the active phase it  
is concentrated near $N\!=\!L$.
 
Of particular interest are the QS scaling properties at the critical
point.  We have verified that $\langle N \rangle \sim L^{1/2}$
in the critical QS state.  The QS lifetime scales in the same manner.
These system-size dependences were encountered previously for
the CP and MVP in the one-site approximation \cite{dickvid}.
The relaxation time $\tau_R$ also grows $ \sim L^{1/2}$ at the
critical point;
we find $\tau_{QS}/\tau_R \sim 2.67$ for $p_1 \!=\! p_2 \!=\! 1/2$.

In the active phase, however, $\tau_{QS}$ grows 
$\sim \exp [{\mbox const.} \times (p_1 \!-\! p_{1c}) L]$,
while $\tau_R$ varies only slightly with $p_1$, $p_2$, and $L$. 
This leads to a clear separation of time scales 
($\tau_{QS} \gg \tau_R$) for large systems.
(For $L\!=\!100$, $p_1\!=\!0.6$, and $p_2\!=\!0.5$, for example,
$\tau_{QS} \!\simeq \! 2000$ while $\tau_R \! \simeq \! 8$.)

In Fig. 5 we show the QS density {\it versus} $p_1$  
in the site approximation, for several system sizes, showing convergence to 
the deterministic mean-field prediction.
Also shown is the moment  
ratio $m$ {\it versus} $p_1$ for the same system sizes. 
Data for $L$ = 1000 - 10$^5$ (see Fig. 6) 
indicate that as $L \to \infty$ at the critical point,  
the moment ratio approaches the value 
1.660, found for the CP in the one-site approximation, and for the 
Malthus-Verhulst process \cite{dickvid}.  

The moment ratio $m$ appears
to approach the same limiting value all along the critical line,
for $p_2 < 1$.   This suggests that the critical QS distribution
has a scaling limit for large $L$, of the form
\begin{equation}
\overline{p}(N) \simeq \frac{1}{\langle N \rangle} 
{\cal P} (N/\langle N \rangle) ~,
\label{scqss}
\end{equation}
where ${\cal P}$ is a scaling function. Fig. 7 
compares (for $p_1 = p_2 = 0.5$), 
$\langle N \rangle \; \overline{p}(N) $ as a function of
$N/\langle N \rangle$, for system sizes $L$= 10$^3$, 10$^4$,
and 10$^5$, as well as the exact scaling function for the
CP and MVP found in Ref. \cite{dickvid}.
It is interesting to note that the QS distribution for the DKCA has the
same scaling form as for the CP and the MVP, despite the fact that
in the critical DKCA, $\overline{p}(N)$ takes its maximum value 
for $N > 1$.  The position of this maximum, however, grows very
slowly with $L$ (roughly, $\sim \ln L$), so that it does not alter the
infinite-size limit. A possible explanation for a maximum away from $N=1$ is 
that, in the DKCA, there are transitions to the absorbing state ($N=0$), from
various values of $N$, not only for $N=1$, as is the case in the CP.

The inset of Fig. 7 shows
$\langle N \rangle \; \overline{p}(N) $ versus $N /\langle N \rangle $
for $p_1 = 0.5$, $L = 2000$, and various values of $p_2$.  The data
collapse confirms the scaling hypothesis, except for $p_2 \simeq 1$. 
The distinct behavior in the latter case is expected,
since $p_2 \!=\! 1$ corresponds to
{\it compact} directed percolation, which has {\it two}
absorbing states (all 0 or all 1).  (The situation is analogous to
that found in the {\it voter model}, a continuous-time process
with two absorbing states \cite{dickvid}.)

We find that any initial distribution not concentrated on $N\!=\!0$
evolves to the QS distribution, which is independent of the 
initial condition.
(Uniqueness of the QS distribution is to be expected in the case of
the DKCA, which has only one absorbing and one active state.
A nonunique QS distribution can be envisioned for a process 
in which the active state exhibits symmetry breaking.)

The direct and iterative methods discussed above yield (as they must),
the same QS distribution.  Using $a$ in the range -0.4 to 0, 
we find that the iterative method converges to the QS distribution 
slightly faster than the direct approach (it typically requires 
about 30\% fewer steps).  In contrast with the continuous-time case, 
in which the iterative method can be orders of magnitude
faster than integration of the master equation \cite{dickvid,intme},
here the gain in efficiency is quite modest.  This is not surprising, since 
the enormous gain in efficiency for continuous-time processes is
associated with the small time step required to maintain numerical
stability, in the usual direct integration schemes.  In the present case,
the direct method has an effective time step of unity.

\section{Pair approximation} 
 
\subsection{Method}

In this section we  
construct the quasi-stationary probability distribution for the 
DKCA at the pair level. 
The system is described by two stochastic variables,  
the number of occupied sites $N$, and the  
number of doubly occupied nearest-neighbor (NN) pairs, $Z$.
We consider a ring of $L$ sites.  [For convenience we introduce
a different notation for the site variables, defining
$\varphi_i = \sigma_{i/2}$ for even $t$ and
$\varphi_i = \sigma_{(i+1)/2}$ for odd $t$.  In this way the
site index always takes the values 1,...,$L$ at all times, and NN sites have
state variables $\varphi_i$ and $\varphi_{i+1}$.]
 
To begin, we establish the allowed range of values for $Z$. Using `1' and  
`0' to represent occupied and vacant sites, respectively, 
we denote by $K$ the number of (10) NN pairs. 
(By symmetry, the number 
of (01) pairs is also $K$.) $K$ is not an independent variable, 
since each 1 is followed by a 0 or another 1, yielding $N=Z\!+\!K$. 
Similarly, the number of (00) pairs, $V$, is given by $V=L-2N+Z$. The  
conditions $K \geq 0$ and $V \geq 0$
imply certain limits for $Z$ on a ring of $L$ sites, listed 
in Table II. 

Next we construct the transition probabilities $W(N,Z|N',Z')$.
Note that the presence of $V'$ (00) pairs at time $t$ implies that
there are at least this many vacant {\it sites} at time $t\!+\!1$;
thus $W \!=\! 0$ for $N > 2N' \!-\! Z'$.  We proceed by analogy with
the one-site approximation: given $N'$ and $Z'$, we first determine
the {\it pair densities} $z = Z/L$, $k$, and $v$ using Eqs.
(\ref{onzez})- (\ref{vdens}).  
(Here $z$, $k$, and $v$ represent the densities at time $t\!+\!1$,
while the variables appearing on the r.h.s. of each equation are
evaluated using $\rho_t = N'/L$ and $z_t = Z'/L$.)
We treat all configurations
having the same $N$ and $Z$ as equally probable, and estimate the
probability of any one such configuration as:
\begin{equation}
Q(N,Z;\rho,z) \equiv \left[ \frac{z^Z k^{2K} v^V}
{\rho^N (1\!-\!\rho)^{L-N}} \right]
\frac{1}{L} \sum_{j=1}^L \frac{s_j s_{j+1}}{p_j},
\label{estq}
\end{equation}
where $k \!=\! \rho \!+\! z$.  The first factor (in square brackets)
is the product of all pair probabilities, divided by the product of
all site probabilitites.  The second factor represents a correction,
needed for normalization of $Q$, which arises as follows.  Suppose we
construct the probability $Q$ starting at site $j$, so that the
first factor in the product is $p_j$, i.e., the pair probability
associated with sites $j$ and $j\!+\!1$.  The next factor will
then be $p_{j+1}/s_{j+1}$, which represents the conditional probability
of the variable at site $j\!+\!2$, given the state of site 
$j\!+\!1$, and so on.
When we close the ring, adding the link between
sites $j\!-\!1$ and $j$, the final factor is
$P(\varphi_{j-1},\varphi_j|\varphi_{j-1},\varphi_j) = 1$, {\it not} 
$P(\varphi_{j-1},\varphi_j)/[P(\varphi_{j-1})P(\varphi_j)]$.  
So the first factor in Eq. (\ref{estq})
has one pair factor,
and two site factors, too many, and should be multiplied by
$s_{j-1}s_j/p_{j-1}$.
Since the position of the starting
link is arbitrary, we take the mean of this correction
over the ring.  

Note that the correction factor may be written so:
\begin{equation}
\frac{1}{L} \sum_{j=1}^L \frac{s_j s_{j+1}}{p_j}
= \frac{1}{L} \left[ Z \frac{\rho^2}{z} + 
2K \frac{\rho (1\!-\!\rho)}{k} + V \frac{(1\!-\!\rho)^2}{v} \right] .
\label{cf2}
\end{equation}
In case the pair numbers take their {\it expected values} 
($Z = L z$, etc.), the correction factor is unity.

The transition probability is the product of a configurational 
probability $Q(N,Z;\rho,z)$ and the number of configurations,
$\Gamma(N,Z;L)$, having exactly
$N$ active sites and $Z$ active pairs on a ring of $L$ sites:
\begl{Wnz} 
W(N,Z|N',Z') = \Gamma(N,Z;L) Q(N,Z;\rho,z) ~,
\fim 
for $(N,Z) \neq (N',Z')$ and $N \leq 2N' \!-\! Z'$; for $ N > 2N' \!-\! Z'$,
$W \!=\! 0$; if $(N,Z) = (N',Z')$,
\begl{Wnza} 
W(N,Z|N',Z') = 1 - \sum_{N,Z}~\!\!\!^* \; \Gamma(N,Z;L) Q(N,Z;\rho,z) ~,
\fim 
where $(^*)$ denotes the exclusion of the single term
$N\!=\!N'$, $Z\!=\!Z'$.
An expression for the combinatorial factor $\Gamma(N,Z;L)$ is
derived in the Appendix.  (Note that $\Gamma \!=\! 0$ for
values of $N$ and $Z$ outside the permitted range given in Table II.)
The evolution of the probability distribution follows:
\begin{equation}
p_{t+1}(N,Z) = \sum_{N',Z'| 2N' -Z' \geq N} W(N,Z|N',Z') p_t(N',Z') ~.
\label{ev2site}
\end{equation}

\subsection{Results}

We constructed the QS distribution at the pair level for the DKCA,
for systems of up to 200 sites, focusing on the behavior in
the vicinity of the critical line.  (The computation is considerably
more demanding of memory and cpu time than is the one-site approximation;
the chief limitation is the evaluation of the coefficients $\Gamma$.)

Fig. 8 shows the QS order parameter {\it versus} $p_1$  
in the pair approximation, for several system sizes. 
We also plot the moment  
ratio $m$, 
showing a series of crossings whose location approaches the critical point 
as $L \to \infty$. The behavior of $m$ {\it versus} $L$, at criticality, 
is shown in Fig 6.
In Fig. 9 we show the QS distribution, $\overline{p}(N,Z)$, at criticality 
($p_2=0.5, \! p_1=0.6306$), for $L=100$. The marginal distribution 
$\overline{p}(N)$ is similar to that found in the site approximation.
The behavior of the mean population, moment ratio, decay rate, 
skewness and kurtosis, as functions of time,   
is again qualitatively similar to that observed in the site
approximation.

\section{Discussion}

We studied the quasi-stationary properties of the DKCA in the one-
and two-site approximations.  Our study represents and extension of
QS analysis, applied to continuous-time models exhibiting an
absorbing-state phase transition in Ref. \cite{dickvid}, to
discrete-time processes.

Compared with continuous-time processes, the numerical analysis
of a discrete-time system is simpler, since it involves iteration rather than 
integration of a set of differential equations.   While this is evident at the
one-site level, at higher levels of approximation the advantage is 
tempered by the fact that starting from a given configuration, transitions
to many (or all) other configurations are possible.  The resulting need for
combinatorial factors (e.g., $\Gamma(N,Z;L) $), complicates the analysis.

An interesting result of our study is that the scaling behavior along the
critical line is the same for the continuous-time contact process
(and the closely related Malthus-Verhulst process) as for the
discrete-time DKCA.  In particular, the QS order parameter
decreases $\sim 1/\sqrt{L}$ in both cases, while the QS lifetime
grows $\sim \sqrt{L}$.  While the universality of global scaling
could have been anticipated on the basis of the central limit theorem,
the similarity extends further, to include the detailed form of the
scaling function governing the QS probability distribution and
its associated moments.  Thus the situation is analogous to that
found numerically in studies of absorbing-state phase transitions:
not only critical exponents, but moment ratios of the order
parameter take universal values at the critical point \cite{rdjaff}.
\vspace{2em}

\noindent{\bf {\small ACKNOWLEDGMENTS}}
\vspace{1em}

This work was financially supported by CNPq, Brazil.
\vspace{2cm}


\begin{table}[htb] 
\caption{DKCA transition probabilities} 
\label{table:1} 
\newcommand{\m}{\hphantom{$-$}} 
\newcommand{\cc}[1]{\multicolumn{1}{c}{#1}} 
\renewcommand{\tabcolsep}{1.5pc} 
\renewcommand{\arraystretch}{1.2} 
\begin{tabular}{@{}lllll} 
\hline 
$\sigma_i | \sigma'_{i-1}, \sigma'_{i+1}$       & \cc{1,1} & \cc{1,0} & \cc{0,1} & \cc{0,0}\\ 
\hline 
1               & $p_2$ & $p_1$ & $p_1$ & 0 \\ 
0               & $1-p_2$ & $1-p_1$ & $1-p_1$ & 1 \\ 
\hline 
\end{tabular} 
\end{table} 

\begin{table}[h] 
\caption{Allowed values for $Z$ on a ring.} 
\label{table:3} 
\newcommand{\m}{\hphantom{$-$}} 
\newcommand{\cc}[1]{\multicolumn{1}{c}{#1}} 
\renewcommand{\tabcolsep}{1.5pc} 
\renewcommand{\arraystretch}{1.2} 
\begin{tabular}{@{}lllll} 
\hline 
\cc{$N$}        &\cc{$Z$} \\ 
\hline 
0,1             & 0     \\ 
2,..., $L/2$    & 0,..., $N-1$  \\ 
$L/2,...L-1$    & $2N-L,...,N-1$  \\ 
$L$             & $L$   \\ 
\hline 
\end{tabular} 
\end{table} 

\centerline{\bf Appendix} 
\vspace{1em} 

To evaluate $\Gamma(N,Z;L)$,
the number of configurations on a ring of $L$ sites with exactly $N$
active sites and $Z$ nearest-neighbor pairs of active sites, we observe that
the associated generating function 
$\zeta (x,y;L) = \sum_{N,Z} \Gamma(N,Z;L) x^Z y^N$, can be written
as the partition function for a one-dimensional lattice gas:
\begin{equation} 
\zeta(x,y;L)  =  \sum_{\sigma_1=0}^1 \cdots \sum_{\sigma_L=0}^1 
{\large x}^{ \sum_i \sigma_i \sigma_{i+1}} \;
{\large y}^{ \sum_i \sigma_i }  ~,
\end{equation} 
with $\sigma_{N+1} \equiv \sigma_1$.
(We note in passing that $x = e^{\beta J}$ and $y=e^{\beta \mu}$   
for the lattice gas with nearest-neighbor interaction $J$, chemical
potential $\mu$, and inverse temperature $\beta$.)
The partition function is evaluated using the transfer matrix 
$T(\sigma,\sigma') = x^{\sigma \sigma'}y^{(\sigma +\sigma')/2}$:
\begin{eqnarray}\label{a:2} 
\zeta(x,y;L) & = & {\mbox Tr}~ T^L~\\ 
	       & = & \lambda^L_1 + \lambda^L_2~,  
\end{eqnarray} 
where $\lambda_{1,2}$ are the eigenvalues of $T$:
\begl{a:3} 
\lambda_{1,2} = 1/2 \left( 1+xy \pm \sqrt{(1-xy)^2 + 4y} \right)~. 
\fim 
For $L$ even, we have: 
\begl{a:4} 
(a+b)^L + (a-b)^L = 2\sum_{n=0}^{L/2} 
{\left(\! \begin{array} {c}
 L \\
 2n
\end{array} \! \right) }
a^{2n} b ^{L-2n}~, 
\fim 
so that 
\begl{a:5} 
\lambda_1^L + \lambda_2^L  = 2 \sum_{n=0}^{L/2}   
{\left(\! \begin{array} {c}
 L \\
 2n
\end{array} \! \right) }
\left (  
\frac{1\!+\!xy}{2} \right)^{2n} \left( \frac{(1\!-\!xy)^2 \!+\! 4y}{4} \right)^{L/2-n}~, 
\fim 
leading to 

\begin{equation}
\label{a:6} 
\lambda_1^L + \lambda_2^L   =  2 y^{L/2} 
\sum_{n=0}^{L/2} \sum_{m=0}^{2n} 
\sum_{p=0}^{L/2-n} \sum_{q=0}^{2p}   
{\left(\! \begin{array} {c}
 L \\
 2n
\end{array} \! \right) }
{\left(\! \begin{array} {c}
 2n \\
 m
\end{array} \!\right) }
{\left(\! \begin{array} {c}
 \frac{L}{2} \!-\!n \\
 p
\end{array} \!\right) }
{\left(\! \begin{array} {c}
 2p \\
 q
\end{array} \!\right) }
\frac{(-1)^q(xy)^{m+q}}{(4y)^{n+p}} ~.
\end{equation}
The coefficient of $x^Z y^N$ is:
\begl{a:7}
\Gamma(N,Z;L) = \frac{2}{4^{\frac{L}{2} + Z -N}}
\sum_{n=0}^{L/2} \sum_{m=0}^{2n} 
{\left(\! \begin{array} {c}
 L \\
 2n
\end{array} \!\right) }
{\left(\! \begin{array} {c}
 2n \\
 m
\end{array} \!\right) }
{\left(\! \begin{array} {c}
\frac{L}{2} - n\\
\frac{L}{2} \!-\! n \!+\! Z \!-\! N
  \end{array} \!\right) }
{\left(\! \begin{array} {c}
 L \!+\! 2(Z\!-\! N \!-\!n) \\
 Z- m
\end{array} \!\right) } 
(-1)^{Z-m}   ~.
\fim
The above expression is evaluated numerically.

\newpage
\noindent FIGURE CAPTIONS
\vspace{1em}


\noindent FIG. 1. DKCA phase diagram - simulation results from
Ref. \cite{atman2}.
\vspace{1em}


\noindent FIG. 2. 
Evolution of the probability distribution, conditioned on survival, 
in the one-site approximation. The initial  
distribution is $p_0(N) = \delta_{N,50}$.  
System size $L\!=\!100$/ $p_1 \!=\! p_2 \!=\! 0.5$.  
\vspace{1em}


\noindent FIG. 3. 
Evolution toward the QS state in the one-site
approximation; parameters as in Fig. 2.  Upper left: decay rate $\gamma$; 
right: mean number of active sites.
Lower left: moment ratio $m$; right: skewness and kurtosis.
\vspace{1em}


\noindent FIG. 4. 
QS distribution at several different points in the DKCA phase diagram, 
for $L=100$.
\vspace{1em}


\noindent FIG. 5. 
Quasi-stationary density for the DKCA, in the site approximation, 
for several system sizes. The inset shows moment ratio $m$ versus $p_1$ for  
the same system sizes.
\vspace{1em}


\noindent FIG. 6. Moment ratio $m$ {\it versus} $L^{-1/2}$ 
at the critical point.
$\Box$: one-site approximation,
$p_1\!=\!p_2\!=\!1/2$; $\circ$: pair approximation, 
$p_1\!=\!0.6306\!~,~~\!p_2\!=\!0.5$. 

\vspace{1em}


\noindent FIG. 7. Scaling plot of the QS probability distribution
in the one-site approximation at the critical point ($p_1\!=\!p_2\!=\!1/2$),
for $L$ = 10$^3$, 10$^4$, and 10$^5$ (curves with maxima approaching
the $y$-axis as $L$ increases), and the asymptotic scaling function
for the contact process found in Ref. \cite{dickvid} (with maximum
at $x\!=\!0$). Inset: Scaling plot of the QS distribution as in Fig. 7,
for $L\!=\!2000$, $p_1 \!=\! 1/2$, and $p_2$ = 0.25, 0.5, 0.75, 0.9,
and 0.999.  The first four curves collapse, while the last has a
broader distribution.
\vspace{1em}


\noindent FIG. 8. 
Quasi-stationary active-site density versus $p_1$
for the DKCA in the pair approximation, 
for several system sizes.  The inset shows the moment ratio $m$ for  
the same system sizes.
\vspace{1em}


\noindent FIG. 9. 
QS pair density probability distribution, conditioned on survival, 
at criticality ($p_2=0.5$, $p_1=0.6306$). 
\vspace{1em}

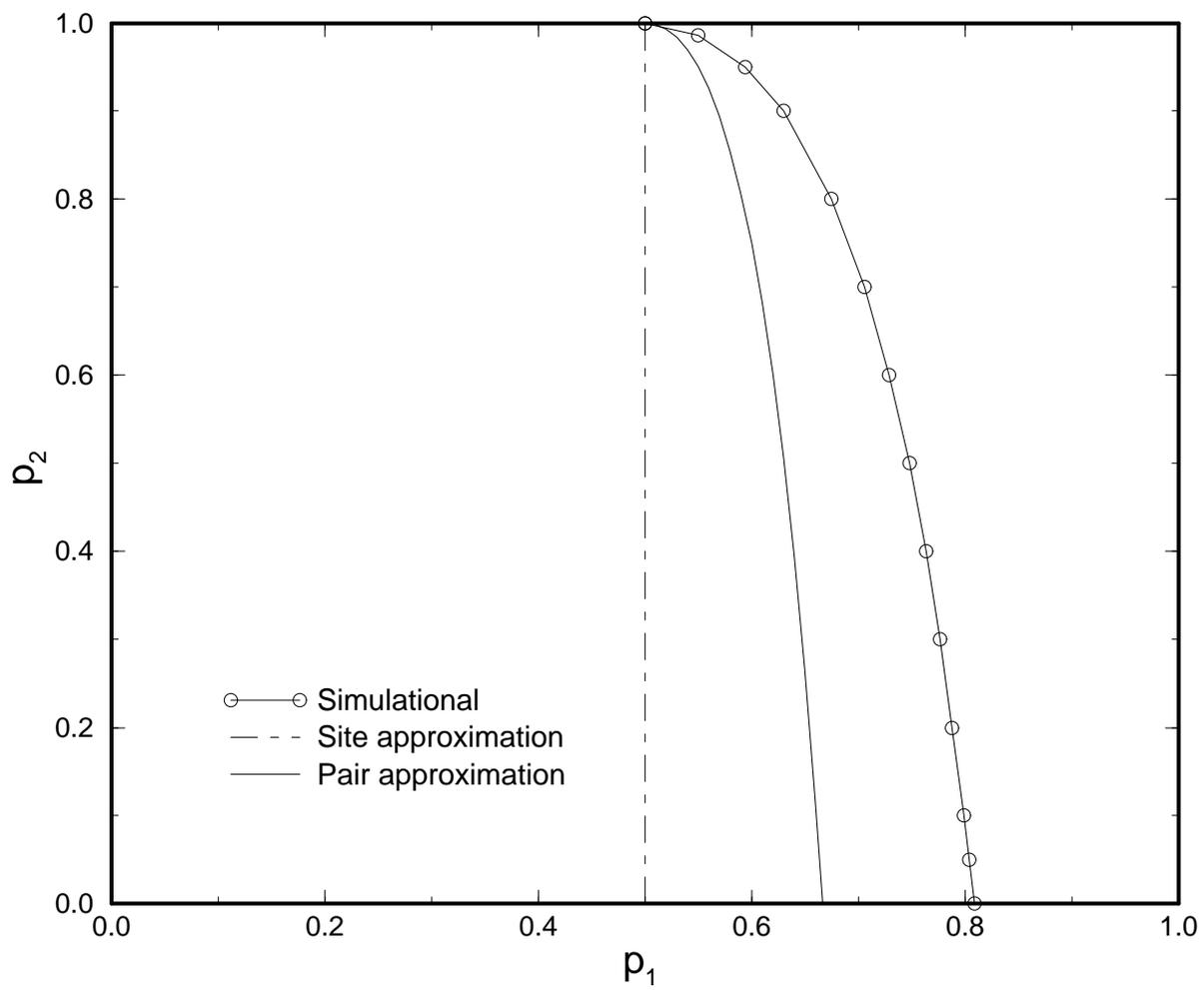
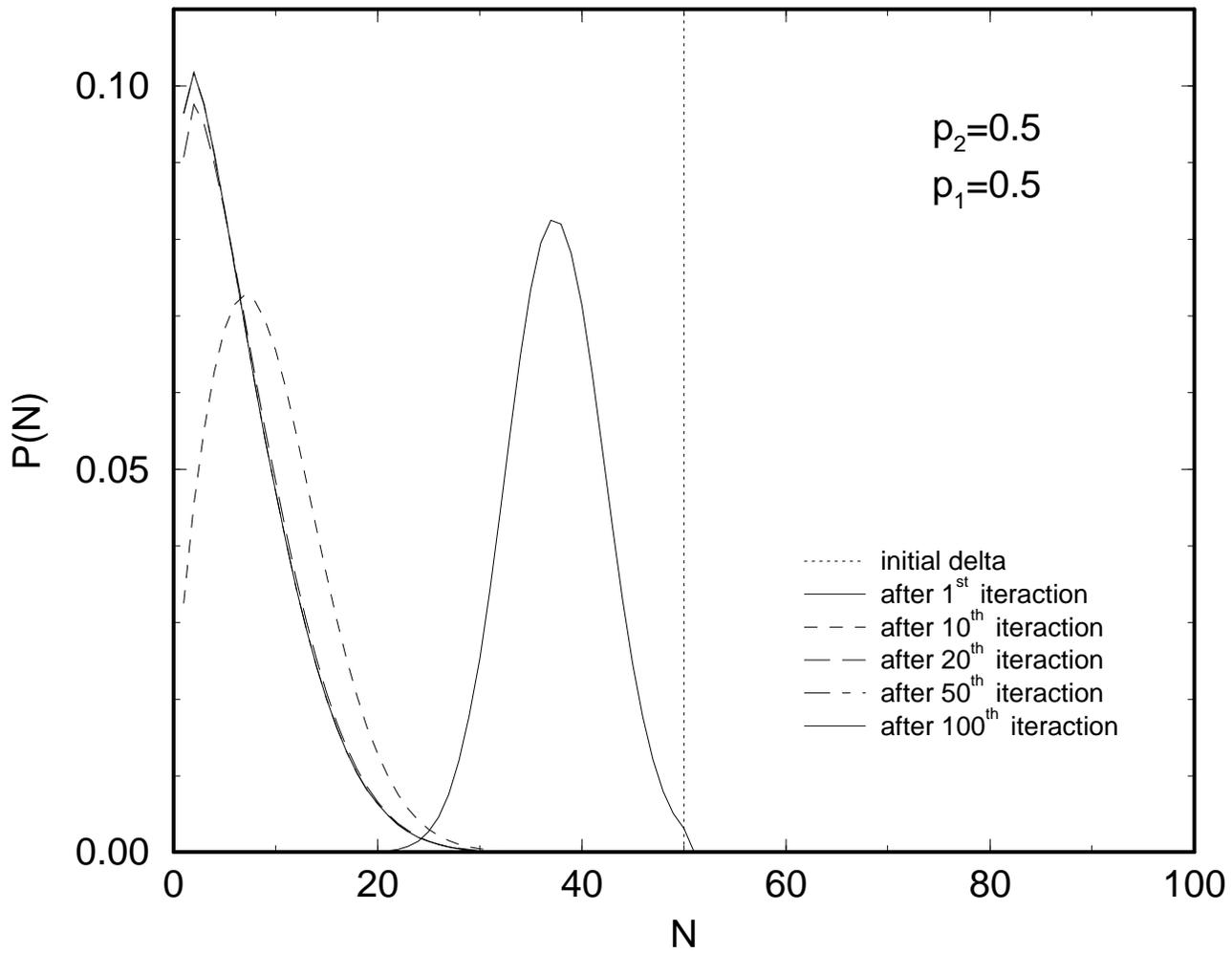
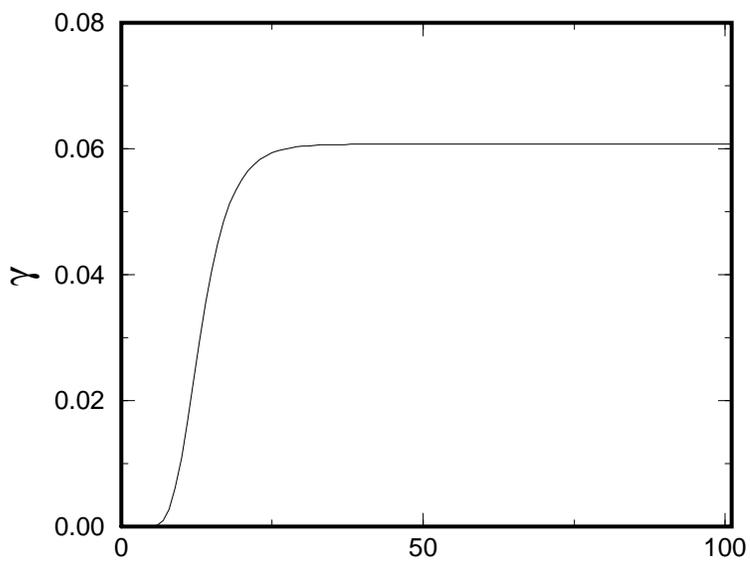
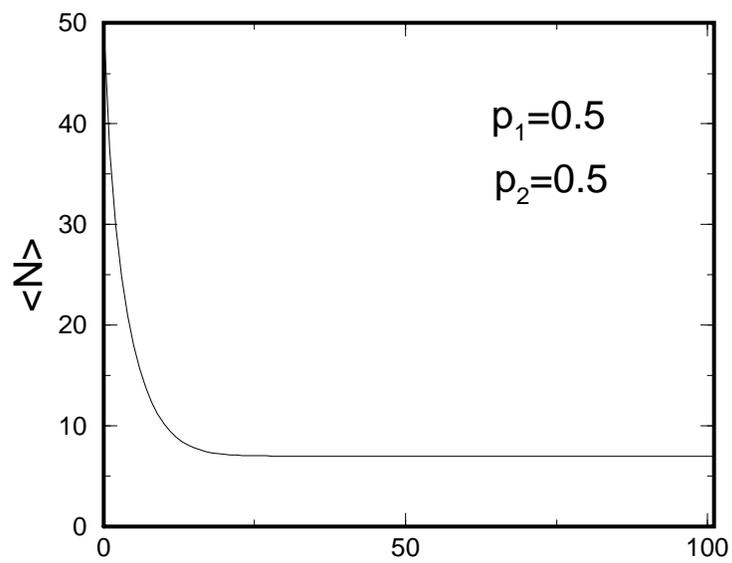
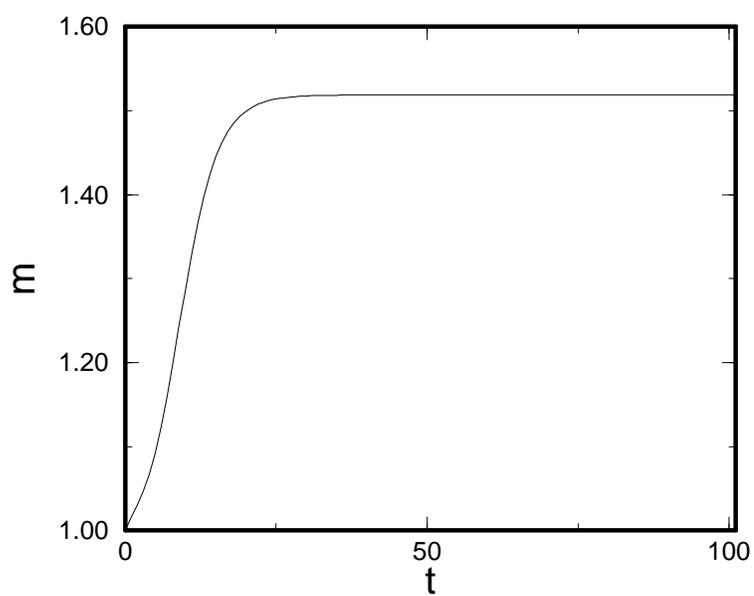
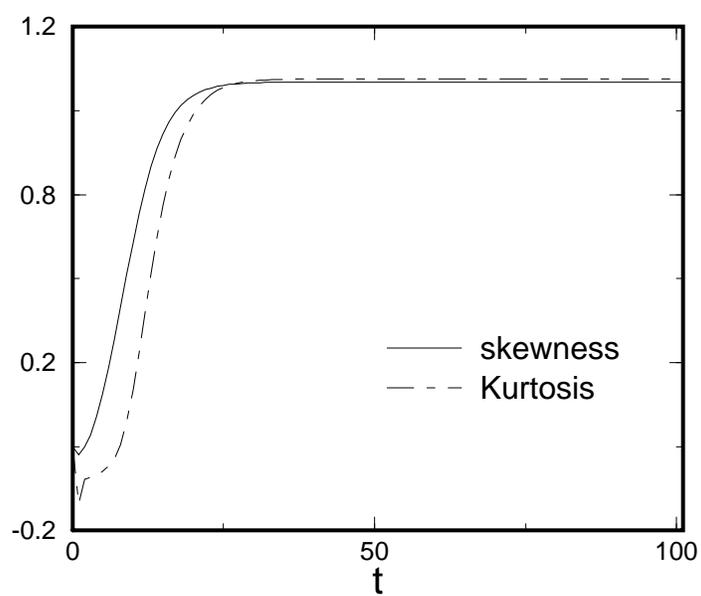
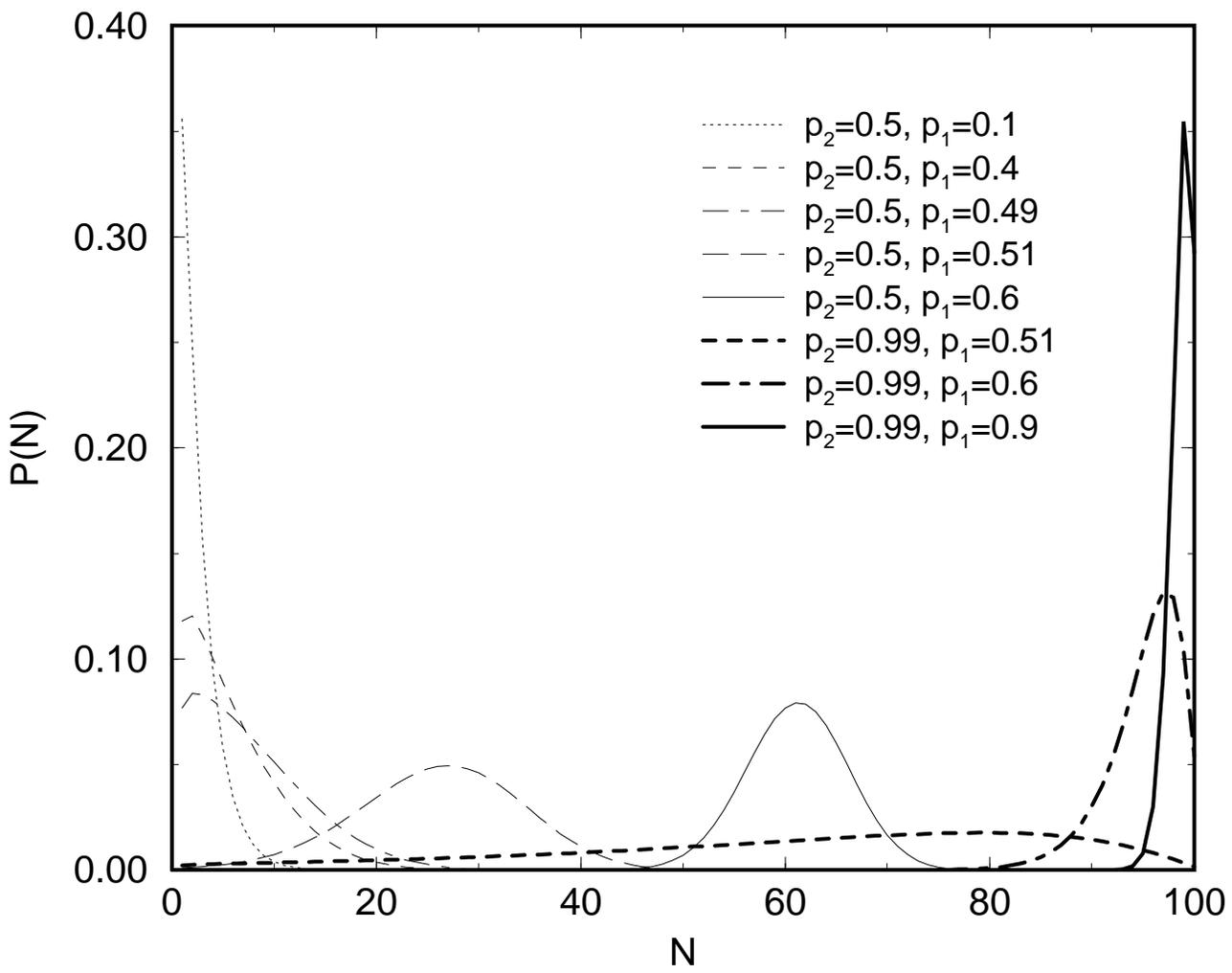
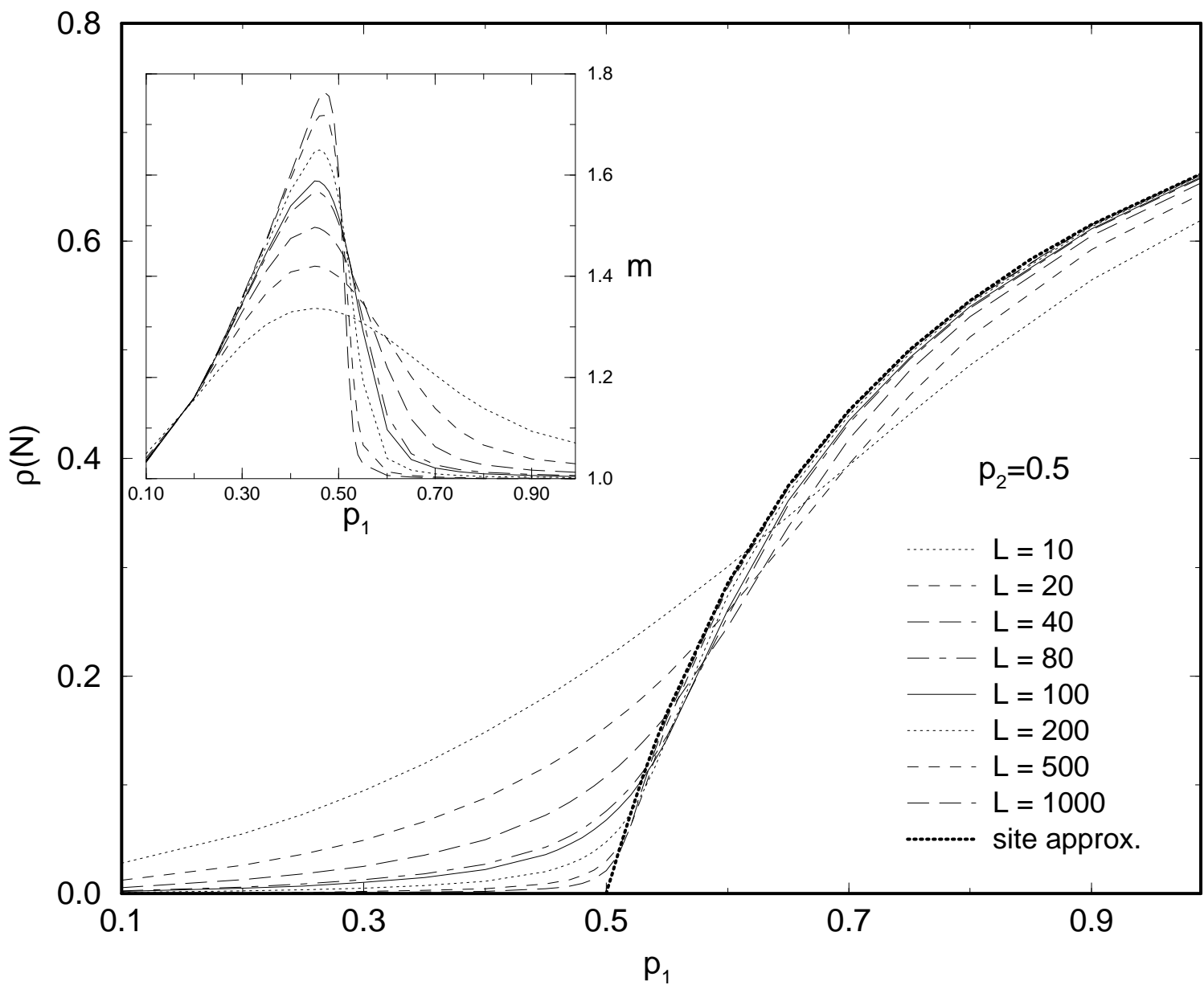
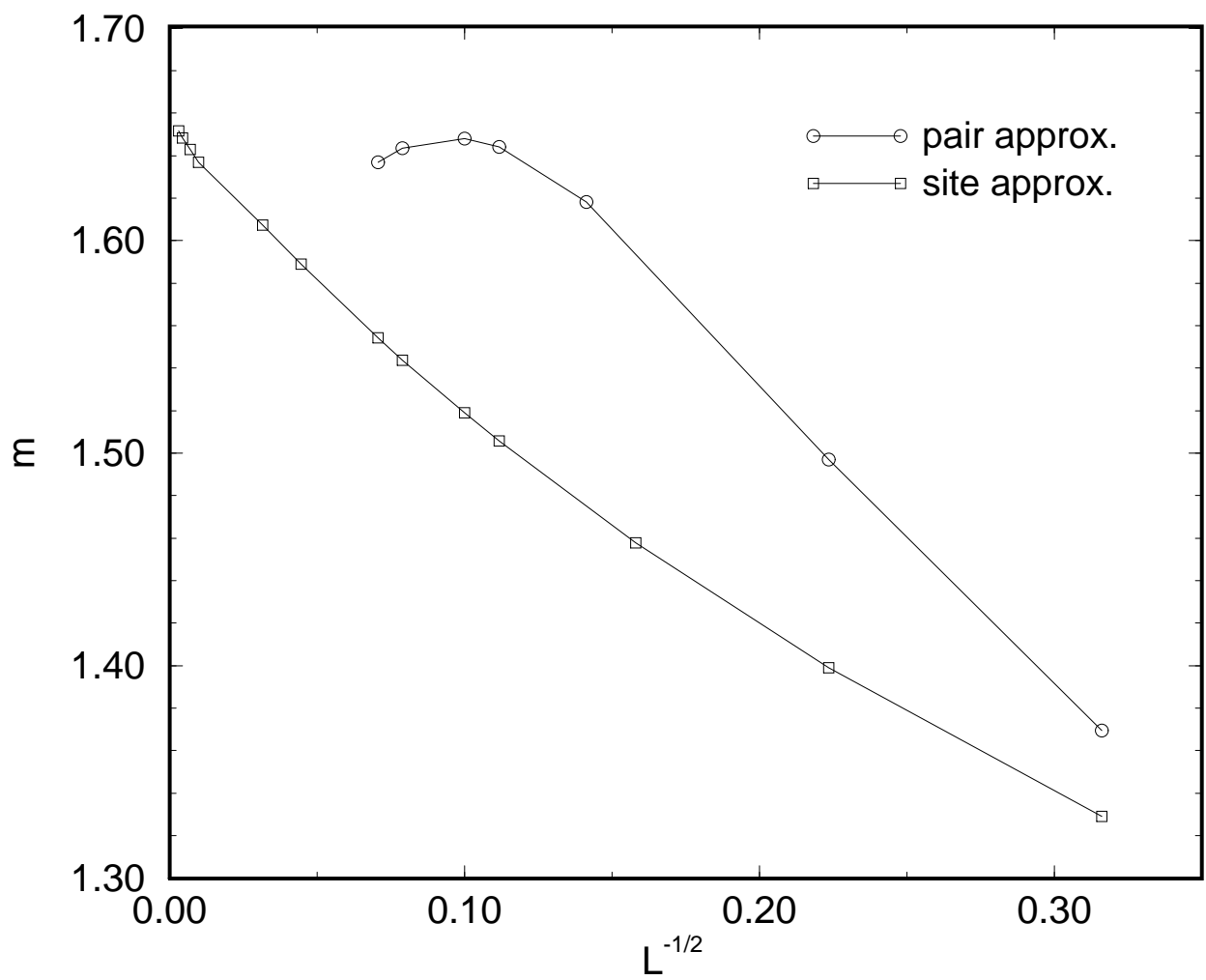
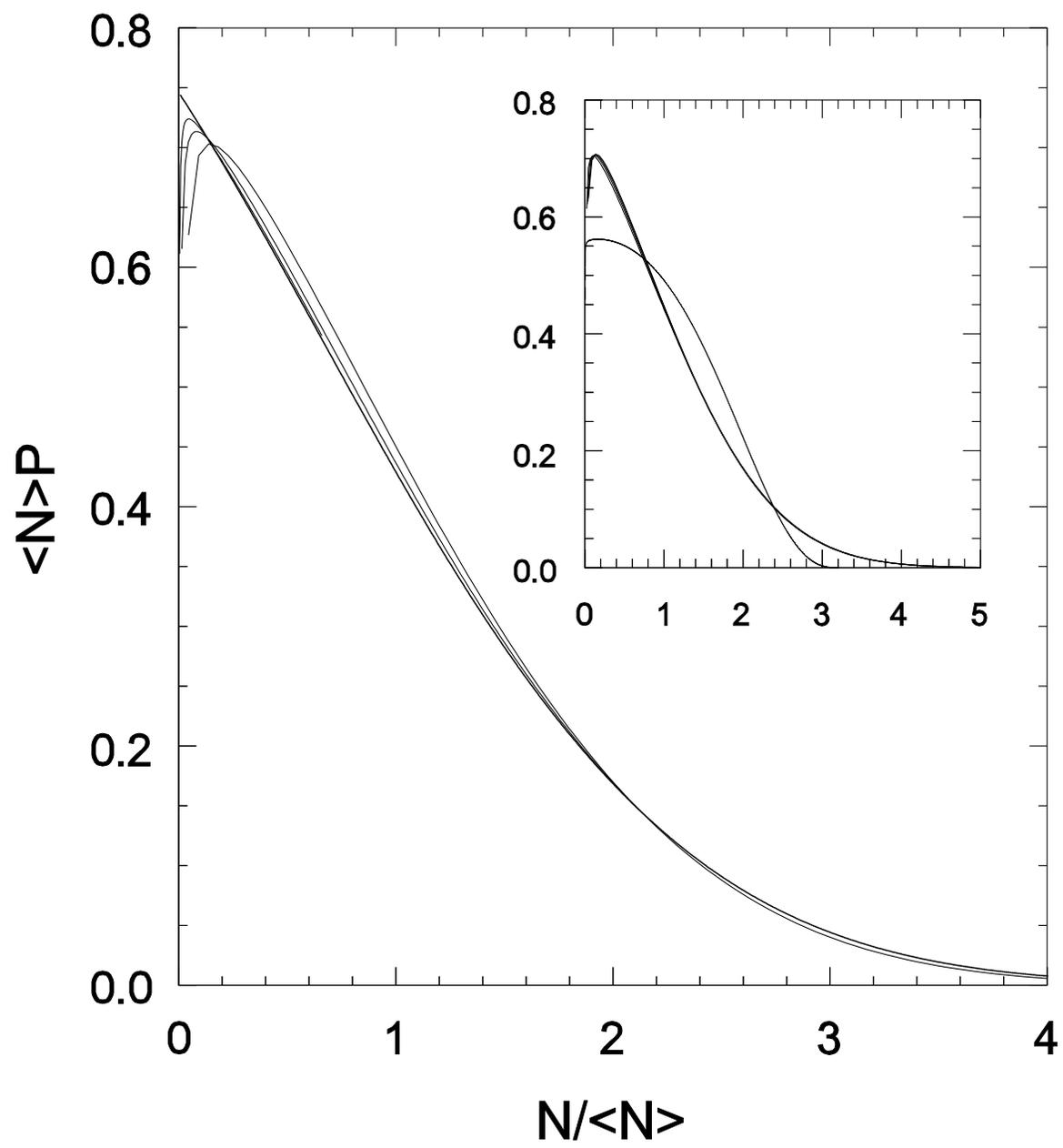
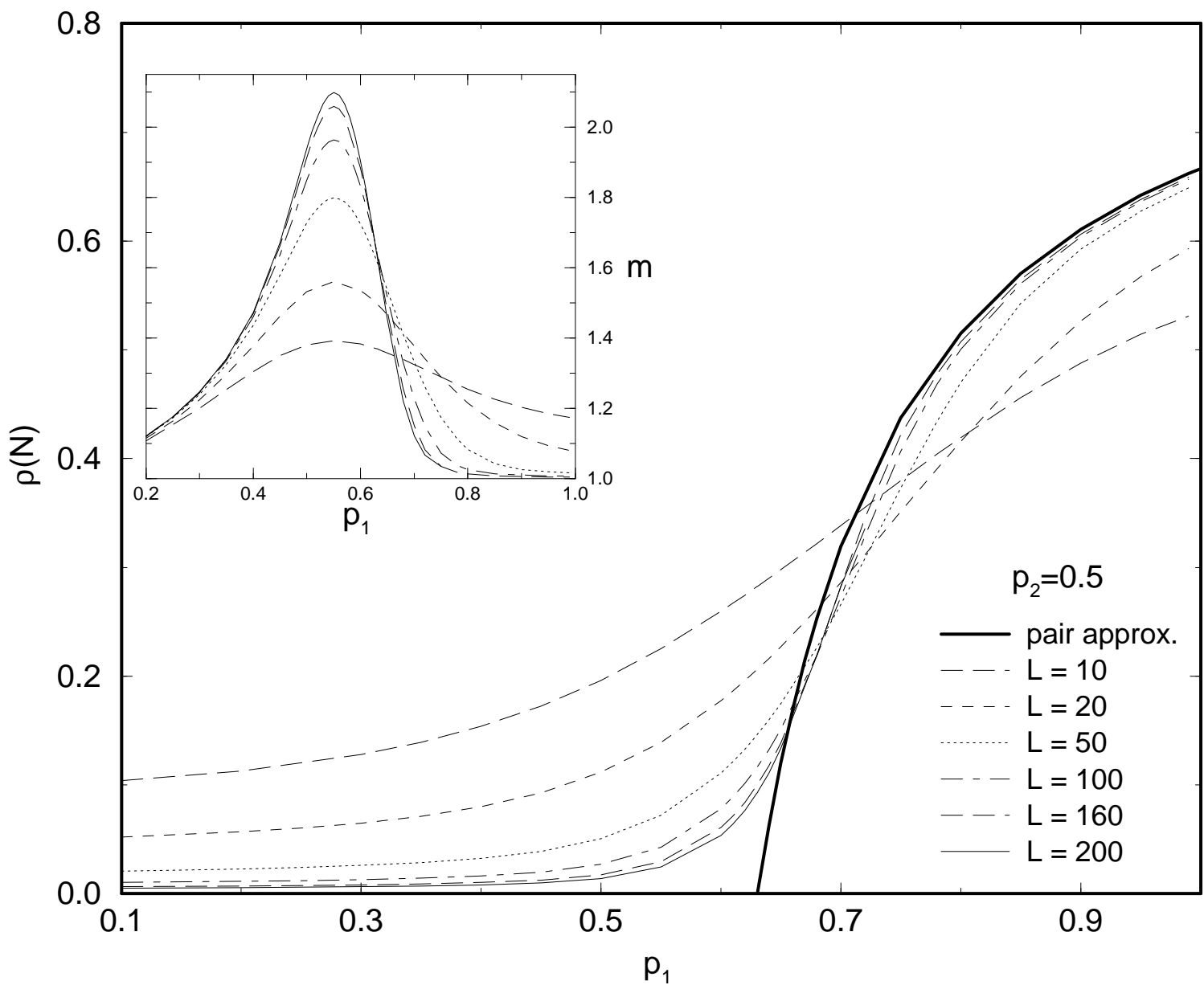
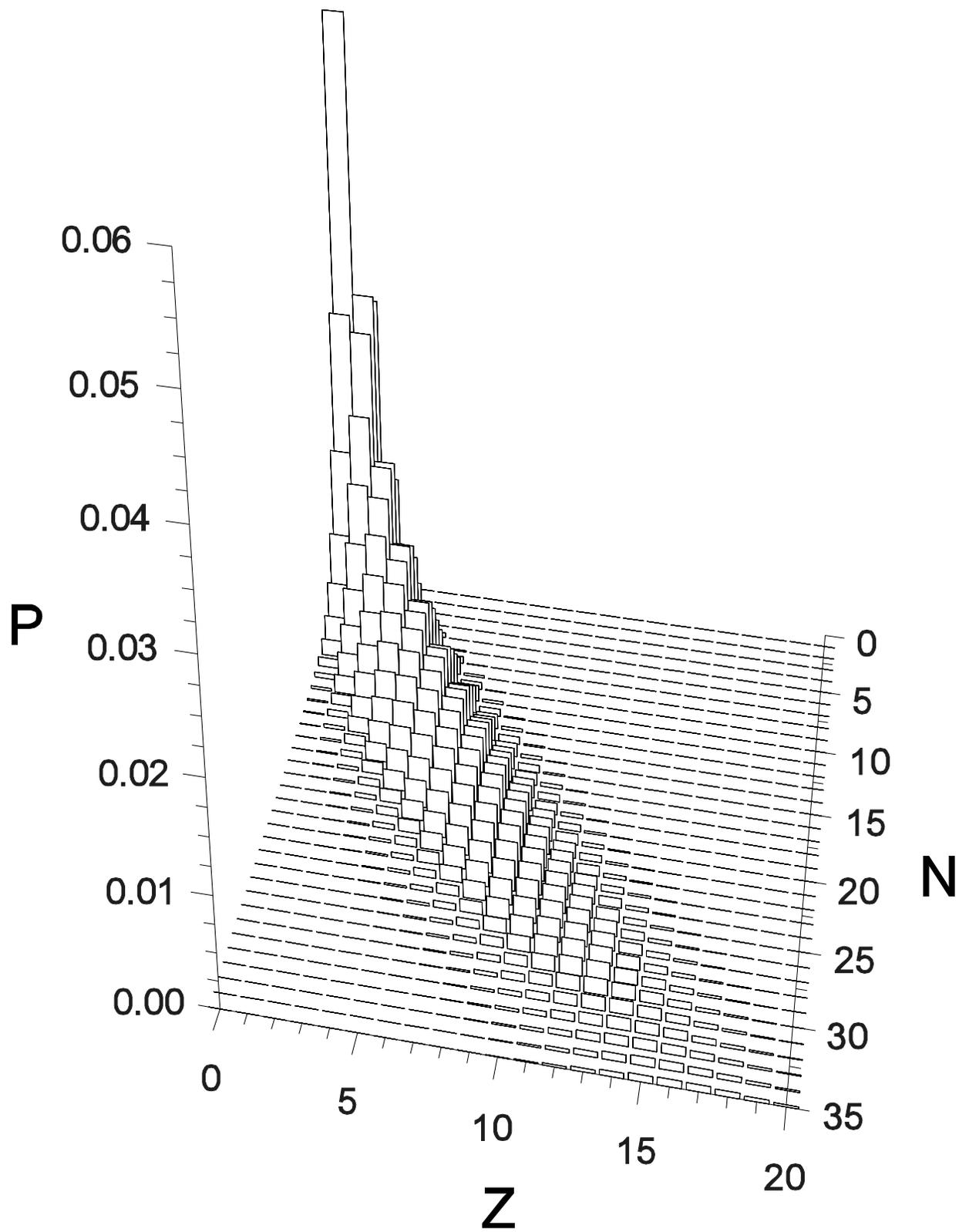
\end{document}